\documentclass[onecollarge]{svjour2}       % onecolumn "king-size"
\usepackage{graphicx}
\usepackage{amssymb}
\usepackage{xcolor}
\begin{document}

\large

\title{Rotation and figure evolution in the creep tide theory. A new approach and application to Mercury}
\author{G. O. Gomes \textperiodcentered \ H. A. Folonier \textperiodcentered \ S. Ferraz-Mello}

\institute{G. O. Gomes and S. Ferraz-Mello \at \emph{Instituto de Astronomia Geof\'isica e Ci\^encias
Atmosf\'ericas, Universidade de S\~ao Paulo, Brasil} \\
 \email{gabrielogomes@usp.br} and \email{sylvio@iag.usp.br}\\}

\titlerunning{}
%\authorrunning{Short form of author list} % if too long for running head

\maketitle
\large
\begin{abstract}
This paper deals with the rotation and figure evolution of a planet near the 3/2 spin-orbit resonance and the exploration of a new formulation of the creep tide theory (Folonier et al. 2018). This new formulation is composed by a system of differential equations for the figure and the rotation of the body simultaneously (which is the same system of equations used in Folonier et al. 2018), different from the original one (Ferraz-Mello, 2013, 2015a) in which rotation and figure were considered separately. The time evolution of the figure of the body is studied for both the 3/2 and 2/1 spin-orbit resonances. Moreover, we provide a method to determine the relaxation factor $\gamma$ of non-rigid homogeneous bodies whose endpoint of rotational evolution from tidal interactions is the 3/2 spin-orbit resonance, provided that (i) an initially faster rotation is assumed and (ii) no permanent components of the flattenings of the body existed at the time of the capture in the 3/2 spin-orbit resonance. The method is applied to Mercury, since it is currently trapped in a 3/2 spin-orbit resonance with its orbital motion and we obtain $4.8 \times 10^{-8}$ s$^{-1} \leq \gamma \leq 4.8 \times 10^{-9}$ s$^{-1}$. The equatorial prolateness and polar oblateness coefficients obtained for Mercury's figure with such range of values of $\gamma$ are the same as the ones given by the Darwin-Kaula model (Matsuyama and Nimmo 2009). However, comparing the values of the flattenings obtained for such range of $\gamma$ with those obtained from MESSENGER's measurements (Perry et al. 2015), we see that the current values for Mercury's equatorial prolateness and polar oblateness are 2-3 orders of magnitude larger than the values given by the tidal theories.
\end{abstract}

\section{Introduction}
\label{sec:intro}

\par It is known that Mercury's current rotation is in a 3/2 spin-orbit resonance with its orbital motion. This means that it rotates on its axis three times for every two revolutions made around the Sun. The scenario used to explain Mercury's current rotational configuration by means of tidal interactions usually involves the assumption that Mercury was initially in a fast rotating state and then evolved under the action of tidal interactions which slowed down the rotation of Mercury until it reached the 3/2 spin-orbit resonance (see e.g Noyelles et al. 2014). The tidal models used to evaluate the torque that slowed down the rotation of Mercury are usually based on classical tidal theories. In these works, the existence of an additional torque arising from a permanent equatorial asymmetry is generally assumed to capture Mercury's rotation in the 3/2 spin-orbit resonance (see e.g Bartuccelli et al. 2017).
\par Recently, Ferraz-Mello (2012, 2013, 2014, 2015a) proposed a theory of tidal interactions based on an approximate solution of the Navier-Stokes equation for a flux that has a low Reynolds number. Considering this theory, the equatorial asymmetry ensues as a consequence of tidal interactions and the capture in spin-orbit resonances can be explained without the assumption of an extra torque due to a permanent equatorial asymmetry of the body. This result was also reported by Makarov and Efroimsky (2013) and confirmed by Correia et al. (2014) with a modelling based on a Maxwell viscoelastic rheology. In the framework of the creep tide theory, tidal interactions alone may be responsible for the evolution of the spin of a non-rigid body and whether or not the body is trapped in a spin-orbit resonance depends on its eccentricity and relaxation factor $\gamma$.
\par In this work, we study the rotational evolution of a homogeneous non-rigid Mercury in the frame of the creep tide theory of Ferraz-Mello (2012, 2013) using the decomposition of the creep equation in three separated equations, as proposed in Folonier et al. (2018). These new equations are virtually equivalent to the equations used in studies employing the Maxwell viscoelastic model (Correia et al., 2014) and they allow for a self-consistent version of the creep tide theory to be constructed. Their simplicity also allows us to use some analytical approximations. We confirm that the capture in spin-orbit resonances depends on $\gamma$, the eccentricity $e$ and the initial value of the rotation rate. The evolution of the equilibrium ellipsoid is discussed for the 3/2 and 2/1 spin-orbit resonances (the synchronous case was already studied in Folonier et al. 2018). We calculate the range of values of the relaxation factor of a non-rigid homogeneous Mercury for which it would currently be in the 3/2 spin-orbit resonance, provided that it was initially in a fast rotating prograde configuration and that its eccentricity minimum value was never below $e=0.1$. We do not make more specific assumptions since there are no observational clues to support them.
\par The report starts with a brief recapitulation of the new formulation of the creep tide theory (Folonier et al. 2018). The differential equations ruling the shape and spin-orbit dynamics in this framework are deduced in Section\,\ref{sec:folonier}. In Section\,\ref{sec:hybrid} we present the solution of such equations corresponding to the approximation where $\dot{ \Omega} \approx 0$ . In Section\,\ref{sec:full}, we study the full model, characterized by the numerical integration of the equations, and use Mercury's physical and orbital parameters to study the rotational dynamics and evolution of its figure. Section\,\ref{sec:application} is dedicated to an application of the equations to the determination of the constraints for the relaxation factor of a body for which the rotational configuration is the 3/2 spin-orbit resonance. We apply our analysis to the case of Mercury's rotation, since this is the only body whose rotation is known to be trapped in a 3/2 spin-orbit resonance. In Section\,\ref{sec:mercury} we discuss Mercury's current figure evolution in the frame of the creep tide theory and compare our results to the ones obtained from the MESSENGER mission (Perry et al. 2015). The conclusions are given in Section\,\ref{sec:conclusion}.

\section{Creep tide theory. Folonier equations}
\label{sec:folonier}

\par Let us consider that an extended body $\tens{m}$ of mass $m$ (primary) experiences the gravitational potential from a point mass body $\tens{M}$ of mass $M$ (companion). The combined action of the centrifugal potential caused by the primary's rotation and the gravitational potential generated by the companion's gravitational attraction causes a deformation on the primary. The equilibrium surface is given by a triaxial ellipsoid (to first order in the flattenings, see Chandrasekhar 1969 and Folonier et al. 2015). In the case of an inviscid fluid (also known as static case), the equilibrium figure of the triaxial ellipsoid is given by
\begin{equation}
\rho (\widehat{\theta},\widehat{\varphi},t) = R \left[ 1 + \frac{1}{2} \epsilon _{\rho} \sin ^2 \widehat{\theta}  \cos (2 \widehat{\varphi} - 2 \varphi ) + \epsilon _z \left( \frac{1}{3} - \cos ^2 \widehat{\theta} \right)  \right] , 
\end{equation}
with $\widehat{\theta}$ and $\widehat{\varphi}$ being the colatitude and longitude of a generic point on the surface of the equilibrium ellipsoid, respectively. $R$ and $\varphi$ are the mean radius of the primary and the true anomaly of the companion. The coefficients $\epsilon _{\rho}$ and $\epsilon _z$ are the equatorial prolateness and polar oblateness of the resulting ellipsoid, respectively, which depend on the physical and orbital parameters of the system through
\begin{equation}
\epsilon _{\rho} = \frac{15M}{4m} \left( \frac{R}{a} \right) ^3 \left( \frac{a}{r} \right) ^3 \equiv \bar{\epsilon} _{\rho} \left( \frac{a}{r} \right) ^3,
\label{nova1}
\end{equation}
\begin{equation}
\epsilon _z = \frac{\epsilon _{\rho}}{2} + \frac{5}{4} \frac{\Omega ^2 R^3}{Gm} ,
\label{nova2}
\end{equation}
with $G$ being the gravitational constant, $\Omega$ the rotation rate. In the general case, where we have a fluid with a given viscosity $\eta$, the equilibrium figure of the resulting ellipsoid is given by $\zeta (\widehat{\theta},\widehat{\varphi},t)$, which is the solution of the creep equation
\begin{equation}
\dot{\zeta} = \gamma (\rho - \zeta), 
\label{creep}
\end{equation}
where the relaxation factor $\gamma$ is inversely proportional to the uniform viscosity coefficient $\eta$ (see Ferraz-Mello 2013 Eq. (3)). Explicitly, we have

\begin{equation}
\gamma = \frac{wR}{2 \eta} ,
\end{equation}
where $w$ is the specific weight on the surface of the body.

\par Folonier et al. (2018) assumed that the solution of the creep equation may be approximated by a generic triaxial ellipsoid with unknown equatorial prolateness, polar oblateness and deviation angle w.r.t the companion (namely, $\mathcal{E} _{\rho}$, $\mathcal{E} _z$ and $\delta$), whose equation is
\begin{equation}
\zeta (\widehat{\theta},\widehat{\varphi},t) = R \left[ 1 + \frac{1}{2} \mathcal{E} _{\rho} \sin ^2 \widehat{\theta}  \cos (2 \widehat{\varphi} - 2 \varphi - 2 \delta ) + \mathcal{E} _z \left( \frac{1}{3} - \cos ^2 \widehat{\theta} \right)  \right] , 
\label{general_ellipsoid}
\end{equation}
\par Substituting Eq.\,(\ref{general_ellipsoid}) into Eq.\,(\ref{creep}) and comparing the terms with the same trigonometrical arguments of $\widehat{\varphi}$ and $\widehat{\theta}$, we obtain three differential equations to be solved, namely
\begin{equation}
\dot{\delta} = \Omega - \dot{\varphi} - \frac{\gamma \epsilon _{\rho}}{2 \mathcal{E} _{\rho}} \sin 2 \delta,
\label{deltadot}
\end{equation}
\begin{equation}
\dot{\mathcal{E}} _{\rho} = \gamma (\epsilon _{\rho} \cos 2 \delta - \mathcal{E} _{\rho}),
\label{dotepsilonrho}
\end{equation}
\begin{equation}
\dot{\mathcal{E}} _{z} = \gamma ( \epsilon _z - \mathcal{E} _ z ).
\label{dotepsilonz}
\end{equation}
\par From the expression for the torque at the companion generated by the primary, we can calculate its reaction ruling the rotational evolution of the primary. The calculations are straightforward (see Folonier et al. 2018). The result is
\begin{equation}
\dot{\Omega} = -\frac{3 G M }{2 r^3} \mathcal{E} _{\rho} \sin 2 \delta .
\label{omegadot2}
\end{equation}
\par Eqs.\,(\ref{deltadot})-(\ref{omegadot2}) rule the evolution of the orientation, shape and rotation of the body. These equations can be integrated numerically to the desired precision.
\subsection{Change of variables}
\par The system of equations presented in the previous section can be simplified if we consider a transformation given by
\begin{equation}
x = E_{\rho} \cos 2 \delta,
\end{equation}
\begin{equation}
y = E_{\rho} \sin 2 \delta,
\end{equation}
where
\begin{equation}
E_{\rho}= \frac{\mathcal{E} _{\rho}}{\bar{\epsilon} _{\rho}} .
\end{equation}
\par The differential equations involving $x$ and $y$ can be easily obtained directly from the above equations. The calculations are straightforward and the resulting differential equations are
\begin{equation}
\dot{x} = \gamma \left( \frac{a}{r} \right) ^3 - \gamma x - y ( 2 \Omega - 2 \dot{\varphi}),
\label{udot}
\end{equation}
\begin{equation}
\dot{y} = - \gamma y + x ( 2 \Omega - 2 \dot{\varphi}),
\label{vdot}
\end{equation}
\begin{equation}
\dot{\Omega} = -\frac{3 G M }{2 r^3} \bar{\epsilon} _{\rho} y .
\label{omegadot}
\end{equation}
\par It is worth mentioning that the above system of differential equations cannot be solved analytically in the general eccentric case. The terms carrying $\dot{\varphi}$ and $(a/r)^3$ are solutions of the Keplerian motion. Another interesting characteristic of the system of differential equations is that the equation for $\mathcal{E} _z$ (see Eq.\,(\ref{dotepsilonz})) is decoupled from the other three differential equations. Thus, it can be treated separately.

\section{Simplified approach with constant rotation}
\label{sec:hybrid}

\par
\begin{figure}
\centering
\includegraphics[height=200pt,width=450pt]{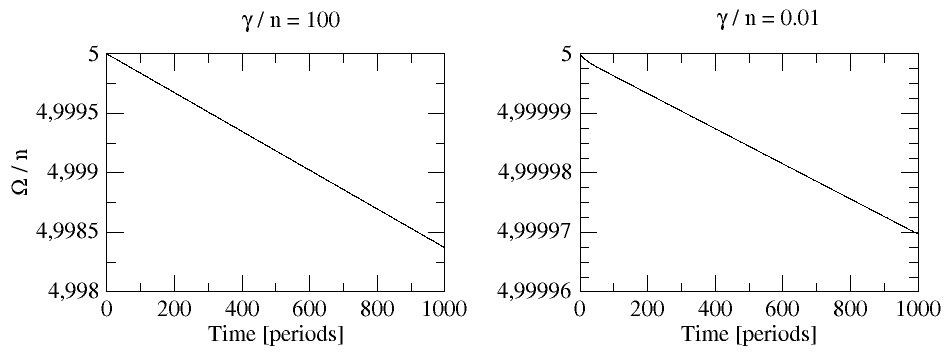}
\caption{Numerical integration of Eqs.\,(\ref{deltadot}) - (\ref{omegadot2}) for a homogeneous non-rigid Mercury considering two different values of $\gamma /n$ as indicated in the figure. The eccentricity is $e=0.2$. In both regimes ($\gamma \gg n$ on the left and $\gamma \ll n$ on the right) we see that the value of the rotation rate changes slowly, in a timescale much larger than the orbital period of Mercury of 88 days.}
\label{fig1}
\end{figure}

\begin{figure}
\centering
\includegraphics[height=250pt,width=370pt]{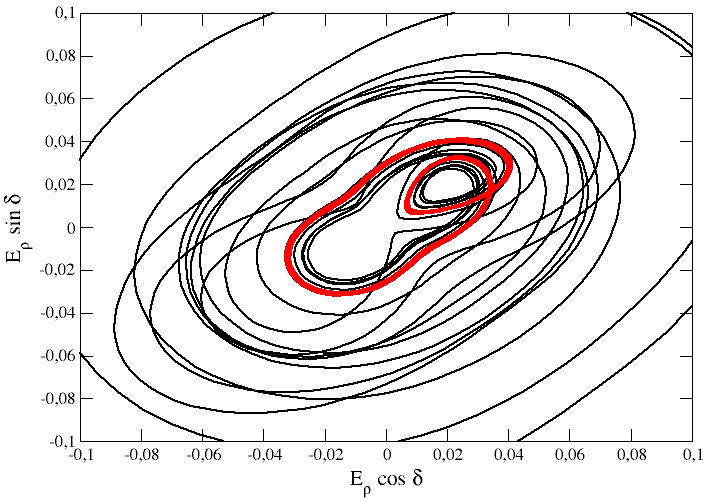}
\caption{Plot of the solutions considering $e=0.15$, $\Omega / n = 2.45$ and $\gamma / n = 4.83 \times 10^{-2}$. After the transient (black curve in the plot), the solution is trapped in a quasi-stationary closed curve (red curve in the plot).}
\label{fig2}
\end{figure}

 In this section we discuss a simplification of the equations of the new formulation of the creep tide theory. In application to Mercury we see in Fig.\,\ref{fig1} that the variation of the rotation rate is small for short time intervals for both regimes of the relaxation factor $\gamma$ ($\gamma \gg n$ and $\gamma \ll n$, where $n$ is the orbital mean motion). Thus, for short time intervals, we can approximate the rotation rate to a constant value. The rotation is rapidly driven to a quasi-steady state. Fig.\,\ref{fig2} shows the solution of the differential equations for $\gamma = 4 \times 10^{-8} \ s^{-1}$. The figure shows both the transient and the quasi-steady state.

\par In order to understand the dynamics associated with the quasi-steady state reached after the transient, we consider initially the simplification in our equations that result when we assume a constant rotation.


\par To solve the system of differential equations for $x$ and $y$, we define the complex variable
\begin{equation}
Z = x + i y,
\end{equation}
and rewrite the equations for $\dot{x}$ and $\dot{y}$ in the simple form
\begin{equation}
\dot{Z} + [\gamma - i(2 \Omega - 2 \dot{\varphi})] Z = \gamma \left( \frac{a}{r} \right) ^3 .
\label{zdot}
\end{equation}
\par The general solution of Eq.\,(\ref{zdot}) is (see Arfken 2005)
\begin{equation}
Z(t) = \frac{\gamma \int e^{\int f(t) dt} \left( \frac{a}{r} \right) ^{3} dt + c}{e^{\int f(t) dt}},
\label{general-z}
\end{equation}
where
$$ f(t) = \gamma - i(2 \Omega - 2 \dot{\varphi}) . $$
\par The integration is straightforward and we use the expansions of the Keplerian motion in Fourier series (see Ferraz-Mello 2015a. Online Supplement):
\begin{equation}
\left( \frac{a}{r} \right) ^{3} e^{2 i \varphi} = \sum _ {k \in \mathbb{Z}} E_{2,2-k} e^{i k \ell} ,
\end{equation}
where $E_{2,j}$ are Cayley coefficients and $\ell$ is the mean anomaly of the companion.
\par The term carrying the constant $c$ in Eq.\,(\ref{general-z}) is transient and it can be neglected. It carries an exponential term and decreases with time tending to zero. The term of $Z(t)$ related to the quasi-steady solution is
\begin{equation}
Z(t) = \gamma \left[ \sum _ {k \in \mathbb{Z}} E_{2,2-k} \int e^{\gamma t + i(k\ell - 2 \widehat{\varphi})} dt \right] e^{-\gamma t + 2 i (\widehat{\varphi} - \varphi)}.
\end{equation}
The calculation is trivial and the resulting expression for $Z(t)$ reads
\begin{equation}
Z(t) = \gamma  \sum _ {k \in \mathbb{Z}} \frac{E_{2,2-k}}{\gamma + i (kn - 2 \Omega)} e^{i(k \ell - 2 \varphi)} .
\end{equation}
\par To obtain $x$ and $y$ we just need to use the identity $Z = x + i y$ and identify the real and imaginary parts of $Z(t)$. The resulting expressions are
\begin{equation}
x(t) = \sum _{k \in \mathbb{Z}} \frac{\gamma E_{2,2-k}}{\gamma ^2 + (2 \Omega - k n)^2} [\gamma \cos (k \ell - 2 \varphi) - (2 \Omega - k n) \sin (k \ell - 2 \varphi)] ,
\label{x_analytic}
\end{equation}
\begin{equation}
y(t) = \sum _{k \in \mathbb{Z}} \frac{\gamma E_{2,2-k}}{\gamma ^2 + (2 \Omega - k n)^2} [\gamma \sin (k \ell - 2 \varphi) + (2 \Omega - k n) \cos (k \ell - 2 \varphi)] .
\label{y_analytic}
\end{equation}

\par From the expressions for $x$ and $y$ thus obtained, we can calculate the corresponding variation in the rotation rate. From Eqs.\,(\ref{omegadot}) and (\ref{y_analytic}), we obtain for $\dot{\Omega}$ that
\begin{equation}
\dot{\Omega} = - \frac{3GM\bar{\epsilon _{\rho}}}{2a^3}  \sum _{k \in \mathbb{Z}} \frac{\gamma E_{2,k}}{\sqrt{\gamma ^2 + (\nu + k n)^2 }} \left\{ \sum _{j \in \mathbb{Z}} E_{2,k+j} \sin \left[ j \ell + \arctan \left( \frac{\nu + k n}{\gamma} \right) \right] \right\} ,
\label{dot_omega_approx}
\end{equation}
where $\nu = 2 \Omega - 2 n$ is the semidiurnal frequency.
\par Eq.\,(\ref{dot_omega_approx}) is exactly the same expression presented in Eq. (36) of Ferraz-Mello (2015a). Thus, we see that the old creep tide theory (Ferraz-Mello 2013, 2015a) and the new version of the theory reproduce the same results if we initially consider $\Omega \approx $const. and then obtain $\Omega (t)$ from the torque expression\footnote{It is worth noting that this equation is also the same equation as Eqn. (39) of Correia et al. (2014) notwithstanding the differences of the Maxwell viscoelastic model adopted by those authors and the creep tide model. This coincidence confirms the fact that the addition of an elastic term to the creep does not change the torque acting on the system, as discussed in Ferraz-Mello (2012, 2013).}. However, we emphasize that the above construction of the solution for constant $\Omega$ is not self-consistent, since we firstly assumed a constant $\Omega$ to obtain $x$ and $y$ and then obtained $\dot{\Omega}$ from the torque expression. As discussed in Folonier et al. (2018), the oscillations in $\Omega$ are significant when studying the energy dissipation in synchronous stiff bodies (see also Efroimsky 2018a, 2018b).

\subsection{Circular case}

\par For the sake of showing in a simpler way some of the consequences of the current simplified approach, we turn our attention to the circular case. In this case, we have $E_{2,0} = 1$ and all the other Cayley coefficientes equal $0$. Then, the solutions for $x$ and $y$ become
\begin{equation}
x(t) = \frac{\gamma ^2}{\gamma ^2 + \nu ^2},
\end{equation}
\begin{equation}
y(t) = \frac{\gamma \nu }{\gamma ^2 + \nu ^2}.
\end{equation}
\par For $\delta$ and $E_{\rho}$, we obtain
\begin{equation}
\delta = \frac{1}{2} \arctan \left( \frac{ \nu }{\gamma} \right) ,
\end{equation}
\begin{equation}
E_{\rho} = \frac{\gamma}{\sqrt{\gamma ^2 + \nu ^2}} = \cos 2 \delta .
\end{equation}
\begin{figure}
\centering
\includegraphics[height=180pt,width=450pt]{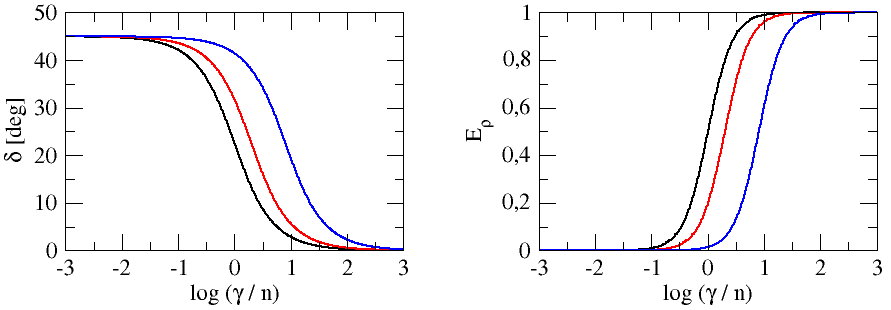}
\caption{Deviation angle $\delta$ (on the left) and normalized equatorial prolateness $E_{\rho}$ (on the right) in the circular case for three different values of $\Omega / n$. We have $\Omega / n = 1.5$, $2$ and $5$ on the black, red and blue curves, respectively.}
\label{fig3}
\end{figure}
\par Fig.\,\ref{fig3} shows the functions $\delta$ and $E_{\rho}$ for the circular case considering different values of $\Omega / n$. We see that for $\gamma \ll n$, the deviation angle $\delta$ is 45$^\circ$ and $E_{\rho}$ is small, of the order $10^{-4}$. As $\gamma$ increases, the value of $\delta$ decreases until it reaches 0$^\circ$. while $E_{\rho}$ increases until $E_{\rho} = 1$. 

\begin{figure}
\centering
\includegraphics[height=300pt,width=420pt]{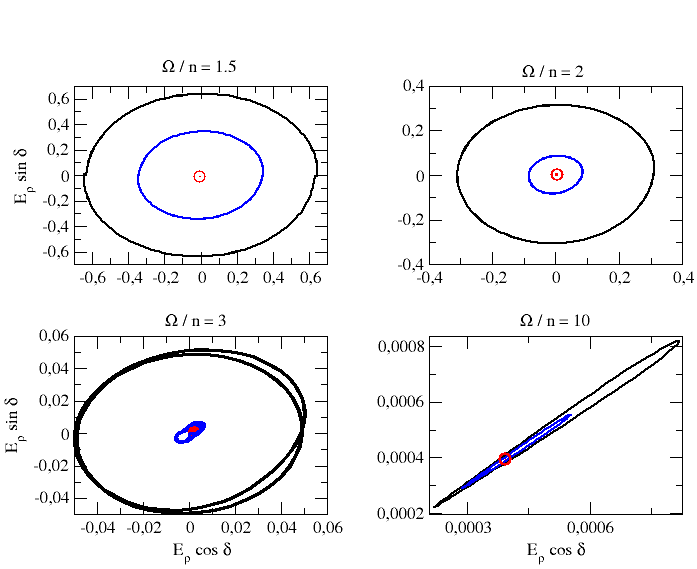}
\caption{Solution of the simplified equations for $\gamma / n = 0.01$ in the circular case (corresponding to the center of the red circle) and two eccentric cases, with $e=0.1$ (blue) and $e=0.2$ (black).}
\label{fig8}
\end{figure}
\par Fig.\,\ref{fig8} shows a comparison of the circular case and the eccentric case. In the circular case, the trajectory in the diagram is given by a single point in each panel. The eccentricity introduces forced oscillations around the circular case whose amplitude increases when the eccentricity increases.

\subsection{Spin-orbit resonances}

\par It was already shown by Makarov and Efroimsky (2013) that, for solid planets and moons (which, in the frame of the creep tide theory, are characterized by a small value of $\gamma$, typically $\gamma  \ll n$), there may be several stationary values of the rotation rate other than the synchronous state. These stationary non-synchronous configurations of the rotation are the spin-orbit resonances.
\par The simplified approach of the new version of the creep tide theory explains how spin-orbit resonances ensue in a given system. For this sake, we calculate the average of $\dot{\Omega}$ (see Eq.\,\ref{dot_omega_approx}) which gives

\begin{equation}
\langle \dot{\Omega} \rangle = - \frac{3GM\bar{\epsilon _{\rho}}}{2 a^3} \sum _{k \in \mathbb{Z}} \frac{\gamma (\nu + k n) E_{2,k} ^2}{{\gamma ^2 + (\nu + k n)^2 }} .  
\label{average-omegadot}
\end{equation}

\par When the rotation is trapped in a spin-orbit resonance, we have $\langle \dot{\Omega} \rangle = 0$. However, in general, the sum of Eq.\,(\ref{average-omegadot}) is not exactly $0$. The equilibrium does not occur at the exact commensurability, but in a neighborhood of it. For a given commensurability, there is a term in the sum of the terms that corresponds to $\nu + kn \approx 0$. We call the index $k$ related to such term as $k'$. Then, the condition for a spin-orbit resonance to be effective becomes
\begin{equation}
\frac{\gamma (\nu + k' n) E_{2,k'} ^2}{\gamma ^2 + (\nu + k' n)^2 } = - \sum _{k \in \mathbb{Z}, k \neq k'} \frac{\gamma (\nu + k n) E_{2,k} ^2}{\gamma ^2 + (\nu + k n)^2 } .
\end{equation} 
\par In the right-hand side of the above equation, we can consider the exact commensurability for all the terms in the sum, that is $\nu = - k' n $. Then, taking into account that the term with $k = k'$ equals $0$, we can write
\begin{equation}
\frac{\gamma (\nu + k' n) E_{2,k'} ^2}{\gamma ^2 + (\nu + k' n)^2 } = - \sum _{k \in \mathbb{Z}} \frac{\gamma (-k'n + k n) E_{2,k} ^2}{\gamma ^2 + (-k'n + k n)^2 } .
\label{final_equality}
\end{equation} 
\par The coefficient of $E_{2,k'}^2$ is of the form $(x+1/x)^{-1}$ and, thus, its maximum value (in absolute value) is attained for $x= \pm1$ or, in this case, when $\gamma = \vert \nu + k' n \vert$, for which the coefficient equals $1/2$. Then, the left-hand side term is never larger than $E_{2,k'} ^2 /2$. As a consequence, the above equation has a solution only if

\begin{equation}
\frac { E_{2,k'} ^2 } {2} \geq \left \vert \sum _{k \in \mathbb{Z}} \frac{\gamma n (k-k') E_{2,k} ^2}{\gamma ^2 + (k-k')^2 n^2 } \right \vert .
\label{resonance_criterion}
\end{equation}

\par For each spin-orbit resonance associated to a given $k'\approx -\nu / n$, the equality on the above equation gives the relation between the minimum eccentricity and the relaxation factor for which the possibility of capture in the given spin-orbit resonance exists (that is, the rotation may be trapped in the resonance as $\Omega / n$ approaches the value corresponding to the resonance). It is important to mention that Eq.\,(\ref{resonance_criterion}) is the same as the one presented in Correia et al. (2014).
\par One application of Eq.\,(\ref{resonance_criterion}) is that we can obtain an analytic relation between the minimum eccentricity for which a given spin-orbit resonance can exist and the relaxation factor $\gamma$. For the 3/2 spin-orbit resonance (corresponding to $k'=-1$), we can expand the Cayley coefficients $E_{2,k}^2$ to order $e^4$ and obtain a quartic equation to be solved in $e$. Since there are only even powers of $e$ (we have $E_{2,k}^2$ only), it can be reduced to a quadratic equation. The only real root gives us the relation between $e_{min}$ and $\gamma$. The calculations are straightforward and the relation for the 3/2 spin-orbit resonance is
\begin{equation}
e_{min}^{(3/2)} = \frac{2\sqrt{2}}{7} \left( \frac{\gamma}{n} \right)^{1/2} - \frac{27 \sqrt{2}}{2401} \left( \frac{\gamma}{n} \right)^{3/2} + \mathcal{O}(\gamma^{5/2}) .
\label{32}
\end{equation}
\par In a similar way, for the 2/1 spin-orbit resonance, we obtain
\begin{equation}
e_{min}^{(2/1)} = \sqrt{\frac{2}{17}} \left( \frac{\gamma}{n} \right)^{1/4} + \frac{59}{51\sqrt{34}} \left( \frac{\gamma}{n} \right)^{3/4} + \mathcal{O}(\gamma^{5/4}) .
\label{21}
\end{equation}
\par It is worth mentioning that the lower the value of $k'$ (corresponding to higher-order spin-orbit resonances), the higher is the order of expansion needed for the Cayley coefficients. For the 7/2 spin-orbit resonance, for instance, we would need at least an expansion of order $e^{10}$.

\section{Complete model. The neighborhood of the resonances}
\label{sec:full}

\par In this section, we study the evolution of a non-rigid homogeneous body for initial values of the rotation rate close to two spin-orbit resonances: the 3/2 and 2/1 spin-orbit resonances.

\subsection{The 3/2 spin-orbit resonance}

\par
\begin{figure}
\centering
\includegraphics[height=300pt,width=400pt]{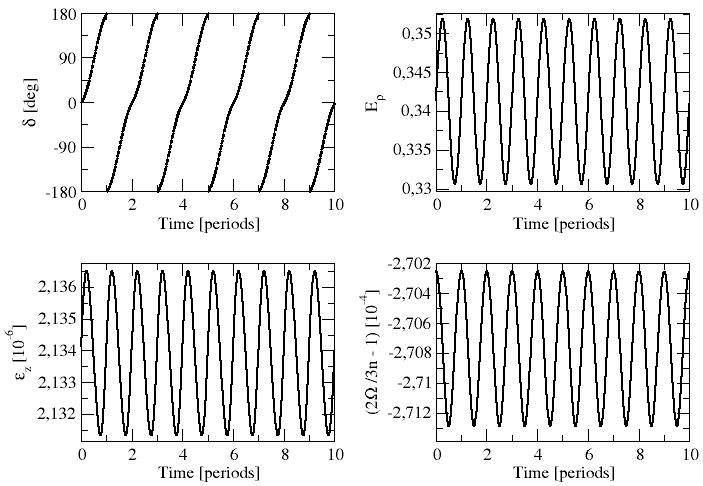}
\caption{Evolution of the lag ($\delta$), flattenings ($E_{\rho}$ and $\mathcal{E}_{z}$) and rotation velocity ($\Omega$) of the resulting ellipsoid for the 3/2 spin-orbit resonance considering a homogeneous non-rigid Mercury with eccentricity $e=0.1$. In this case, we adopted $\gamma / n = 0.01$ (corresponding to $\gamma = 8.27 \times 10^{-9}$ s$^{-1}$).}
\label{fig11}
\end{figure}

 The 3/2 spin-orbit resonance is the last one encountered by a tidally evolving body before it reaches synchronization. Fig.\,\ref{fig11} shows the evolution of the shape, orientation and rotation of a non-rigid homogeneous Mercury when the rotation already evolved and reached the 3/2 spin-orbit resonant configuration. 

\par The numerical results presented in Fig.\,\ref{fig11} show that once the rotation is trapped in the 3/2 spin-orbit resonance, the equatorial prolateness and polar oblateness suffer only small oscillations around a mean value. The same behavior holds for the rotation rate $\Omega / n$. However, for the orientation of the equatorial tidal bulge $\delta$, we observe that the behavior is given by a non-uniform circulation with a period that equals the orbital period of Mercury. Thus, it behaves almost like a rigid body regarding the orientation of the equatorial tidal bulge. This can be seen if we consider Eq.\,(\ref{deltadot}) and make $\gamma = 0$ (corresponding to a rigid body). In this case, the differential equation becomes
\begin{equation}
\dot{\delta} = \Omega - \dot{\varphi} ,
\end{equation}
the solution of which shows that $\delta$ circulates with a period
\begin{equation}
T_{\delta} = \frac{2 \pi}{\Omega - \dot{\varphi}}.
\label{tdelta}
\end{equation}
\par Since we have, in the 3/2 spin-orbit resonance, $\Omega / n \approx 3/2$, it follows that $T_{\delta} = 4 \pi / n$. Thus, in the case of a rigid body in the 3/2 spin-orbit resonance, $\delta$ circulates with a period equal to twice the orbital period of Mercury. Fig.\,\ref{fig11} shows the evolution of the shape, orientation and rotation of the body when $\gamma / n = 0.01$ for the 3/2 spin-orbit resonance. For smaller values of $\gamma / n$, the mean value of $\Omega$ in the 3/2 spin-orbit resonance is higher, and the amplitudes of the oscillations become smaller.

\par
\begin{figure}
\centering
\includegraphics[height=270pt,width=350pt]{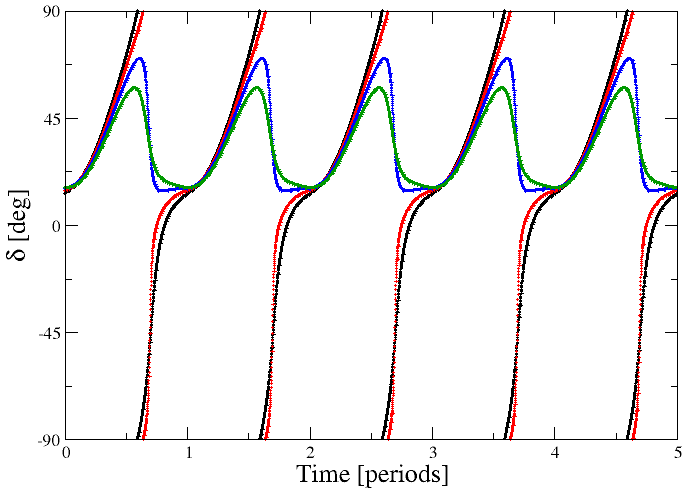}
\caption{Evolution of $\delta$ in the neighborhood of the condition $\Omega / n = 3/2$. The $\gamma / n$ factor varies in each curve in the following way: $\gamma / n = 0.251$, $0.316$, $0.398$ and $0.501$ in the black, red, blue and green curves, respectively. In all cases, the eccentricity is fixed as $e=0.1$. The behavior of $\delta$ depends not only on the value of $\Omega / n$, but also on the ratio $\gamma / n$. We see from the red and blue curves that the limit value for which $\delta$ circulates or oscillates lies between $0.398$ and $0.316$.}
\label{fig14}
\end{figure}

 Fig.\,\ref{fig14} shows the evolution of $\delta$ in the neighborhood of the 3/2 spin-orbit resonance for four different values of $\gamma / n$. The behavior of $\delta$ strongly depends on the value of $\gamma / n$. As $\gamma$ increases, there is a critical point where the regime of $\delta$ changes from circulation to oscillation. The transition between the two regimes can be seen comparing the blue and the red curves in the figure. The results in Fig.\,\ref{fig14} show that the body behaves almost like a rigid body even for values of $\gamma / n$ for which the body is not necessarily trapped in the spin-orbit resonance. It is worth emphasizing that the 3/2 spin-orbit resonance is the case of Mercury's rotational configuration.

\par
\begin{figure}
\centering
\includegraphics[height=250pt,width=250pt]{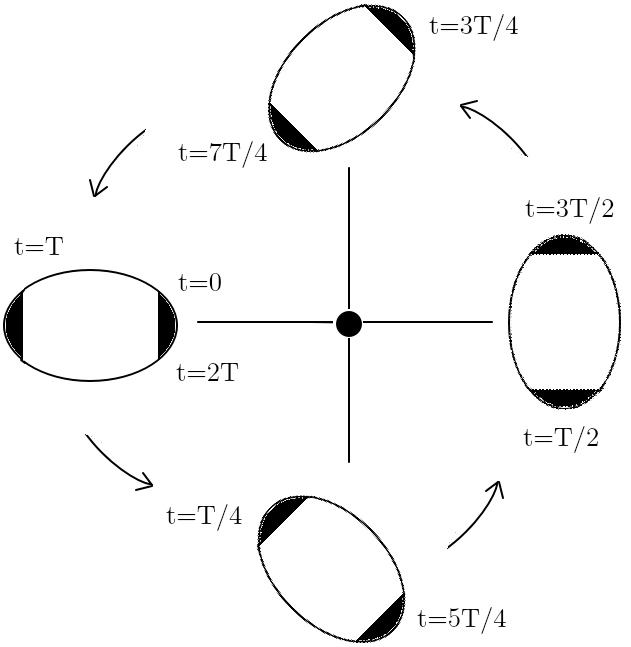}
\caption{Geometrical configuration of the resulting ellipsoid of Mercury in the case of the 3/2 spin-orbit resonance when $\gamma$ is \textbf{very} small. In this case, the body behaves like a rigid body. The configuration at different times is shown along with the respective time on the side of the black areas. After one orbital revolution, the orientation of the ellipsoid is reversed (i.e the orientation of the ellipsoid is rotated by $540$ deg.). The arrows indicate the chronological sequence of the figures.}
\label{fig13}
\end{figure}

 Fig.\,\ref{fig13} shows the geometrical configuration of the ellipsoidal figure of the body, with bulges indicated by black areas for different times. From the figure, it can be seen that it behaves almost like a rigid body when it is trapped in the 3/2 spin-orbit resonance. For every 2 orbital periods, $\delta$ circulates from $0$ to $2\pi$.

\subsection{The 2/1 spin-orbit resonance} 

\par In this subsection, we perform a discussion similar to the one presented in the previous subsection, but for the 2/1 spin-orbit resonance.

\par
\begin{figure}
\centering
\includegraphics[height=300pt,width=400pt]{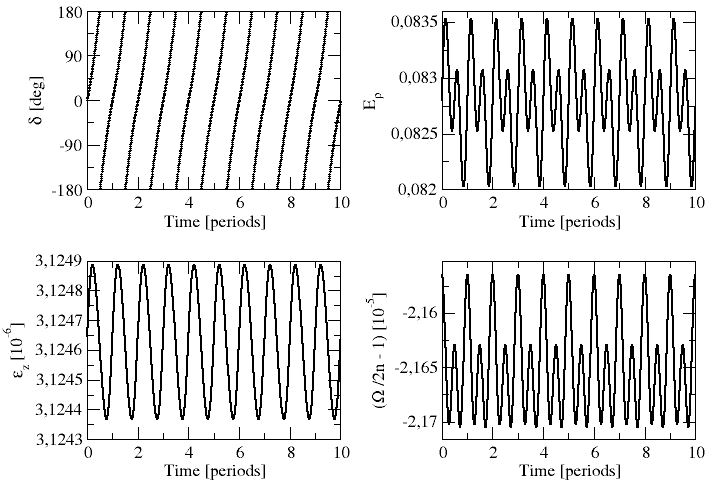}
\caption{Evolution of the lag ($\delta$), flattenings ($E_{\rho}$ and $\mathcal{E}_{z}$) and rotation rate $\Omega$ of a non-rigid body in the 2/1 spin-orbit resonance with $\gamma / n = 0.001$ (corresponding to $\gamma = 8.27 \times 10^{-10}$ s$^{-1}$) and $e=0.1$. The behavior of $E _{\rho}$, $\mathcal{E}_{z}$ and $\Omega / n$ is qualitatively the same as in the 3/2 spin-orbit resonance.}
\label{fig15}
\end{figure}

\begin{figure}
\centering
\includegraphics[height=250pt,width=250pt]{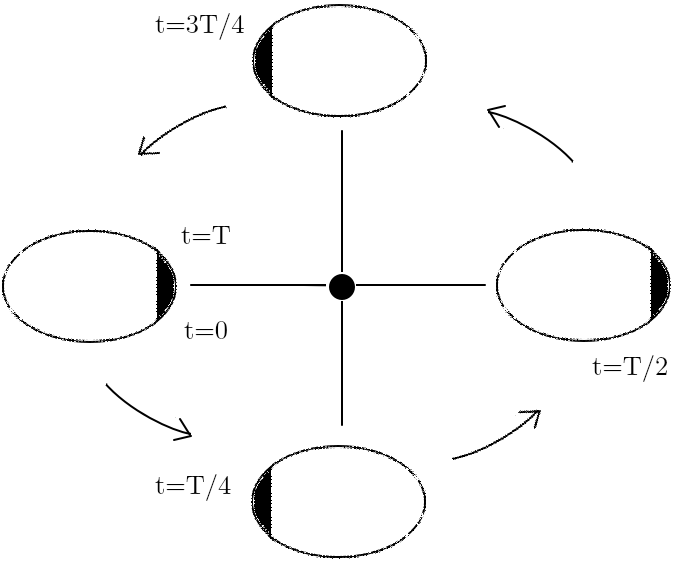}
\caption{Geometrical configuration of the resulting ellipsoid in the case of the 2/1 spin-orbit resonance when $\gamma$ is \textbf{very} small. In this case, the body behaves like a rigid body. As in the Fig.\,\ref{fig13}, the arrows indicate the chronological sequence of the figures.}
\label{fig16}
\end{figure}

 The results in Fig.\,\ref{fig15} show the same qualitative features for the equatorial prolateness, polar oblateness and rotation rate as it was done in the case of the 3/2 spin-orbit resonance. The only additional interesting feature is that for the 2/1 spin-orbit resonance, the oscillations in $E_{\rho}$ and $\Omega / n$ have two noticeable harmonic components. One of the components has a period of $2 \pi / n$, while the second component has a period of $\pi / n$ (cf. top graph on the right in Fig.\,\ref{fig15}). Regarding the behavior of $\delta$, the period of circulation in the case of the 2/1 spin-orbit resonance is equal to the orbital period of Mercury. This result is important because, as in the case of the 3/2 spin-orbit resonance, the body behaves almost like a rigid body when it is trapped in the spin-orbit resonance (also in agreement with Eq.\,(\ref{tdelta})). The geometrical configuration of the body's figure can be seen in Fig.\,\ref{fig16}. The ellipsoidal bulge rotates around the body's center of mass with the same period \textbf{as} the orbital motion w.r.t the position of the companion (which corresponds to $\delta$). If we considered $\varphi _B = \varphi + \delta$, corresponding to an origin in a fixed point in space, the period would be half of the orbital period.

\begin{figure}
\centering
\includegraphics[height=270pt,width=350pt]{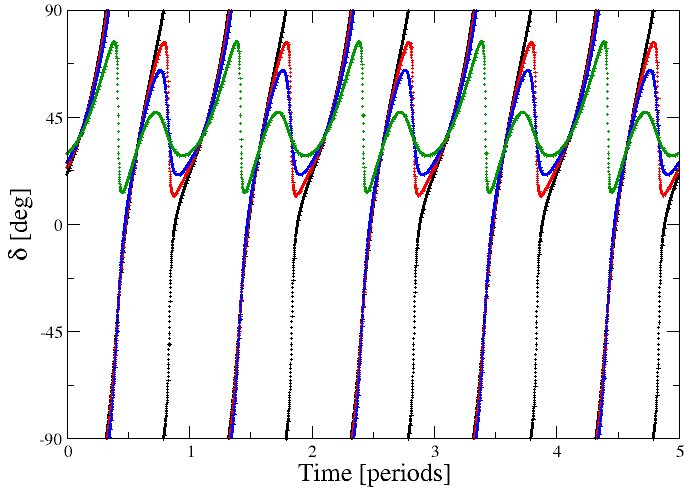}
\caption{Evolution of $\delta$ in the neighborhood of the condition $\Omega / n = 2$. The $\gamma / n$ factor is equal to $0.1$, $0.126$, $0.158$ and $0.3$ in the black, red, blue and green curves, respectively. In the black curve, the period of circulation of $\delta$ is equal to half of the orbital period. In the blue and red curves, the period of circulation increases to one orbital period and in the green curve, $\delta$ oscillates around 45 deg.}
\label{fig17}
\end{figure} 

\par In Fig.\,\ref{fig17}, we repeat the analysis done for the 3/2 spin-orbit resonance regarding the evolution of $\delta$ in the proximity of the resonance. We can see that even for a value of $\gamma / n$ higher than the critical value necessary for the 2/1 spin-orbit to exist, $\delta$ circulates with a period equal to the orbital period (see black curve in Fig.\,\ref{fig17}). Thus, there is a range of values for $\gamma / n$ for which the period of circulation is equal to the orbital period. If $\gamma / n$ is higher than such range of values, the period of circulation becomes twice the orbital period, which was the case for the 3/2 spin-orbit resonance (see red and blue curves in Fig.\,\ref{fig17}). Finally, the green curve in Fig.\,\ref{fig17} shows that $\delta$ oscillates around 45 deg. when $\gamma / n$ is even higher than the range for which $\delta$ circulates with a period equal to twice the orbital period. Therefore, for the 2/1 spin-orbit resonance, there are three possible behaviours for $\delta$ that depend on the value of $\gamma / n$.

\section{Trapping into the 3/2 spin-orbit resonance}
\label{sec:application}

\par In this section we use the same equations as Folonier et al. (2018) to study the rotational evolution of a body that is captured in the 3/2 spin-orbit resonance as a consequence of tidal interactions and analyze the range of values of $\gamma$ able to drive the body's rotation to this spin-orbit configuration.

\begin{figure}
\centering
\includegraphics[height=320pt,width=450pt]{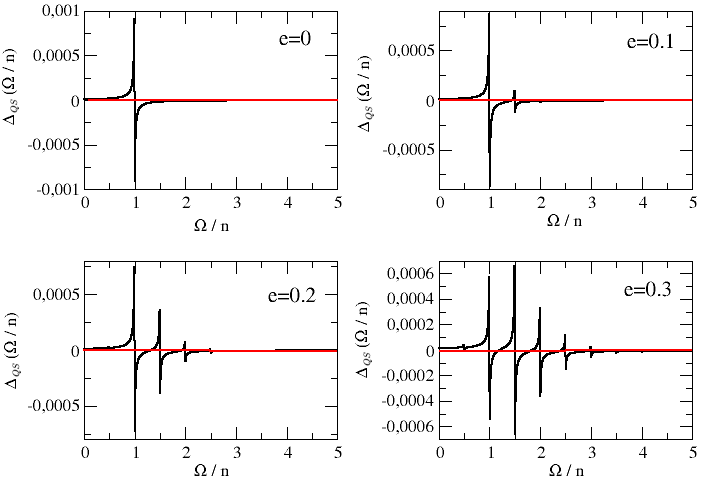}
\caption{Secular variation of the rotation rate $\Delta(\Omega/n)$ as a function of the rotation rate $\Omega$ normalized by the mean-motion when $\gamma/n=10^{-2}$. When the eccentricity increases, the number of visible stable solutions also grows, as can be seen by comparing the panels in the figure.}
\label{fig5}
\end{figure}

\par Fig.\,\ref{fig5} shows the secular variation of the quasi-steady rotation rate $\Delta _{QS} (\Omega / n)$ as a function of the rotation rate $\Omega / n$. To obtain the secular variation of the rotation rate, we integrated Eqs.\,(\ref{deltadot}) - (\ref{omegadot2}) with a given initial rotation rate until the rotation was captured in a quasi-steady state. Then, for each 100 orbital periods, we calculated the difference $\Omega / n$ w.r.t the previous value 100 periods before. Many aspects are presented in Fig.\,\ref{fig5}. We see that, for $\gamma \ll n$, the number of possible spin-orbit resonances is larger for higher eccentricities (compare the panels on Fig.\,\ref{fig5}). The appearance of stable points (i.e possible spin-orbit resonances) for larger eccentricities and low values of $\gamma / n$ was already studied in Correia et al. (2014) and Ferraz-Mello (2015a). The results presented in Fig.\,\ref{fig5} are in agreement with both these works.

\subsection{Constraints for the relaxation factor}

\par In order to evaluate the relaxation factor of a body whose final rotational configuration is the 3/2 spin-orbit resonance, we perform full numerical integrations of Eqs.\,(\ref{deltadot}) -\,(\ref{omegadot2}) for selected initial conditions just above the 3/2 and 2/1 spin-orbit resonances.
\par
\begin{figure}[!h]
\centering
\includegraphics[height=320pt,width=450pt]{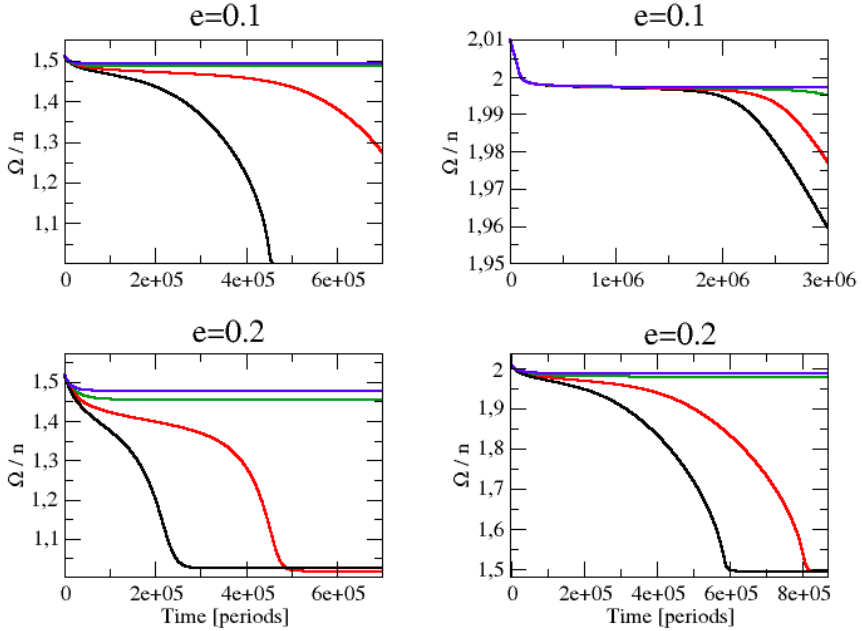}
\caption{Evolution of the rotation rate of the body considering different values of $\gamma$ (varying in uniform steps from the blue to the black curve from the lowest to the highest values) and for two eccentricities. We have $e=0.1$ on the panels on the top and $e=0.2$ on the panels on the bottom. In the top left panel we have $\gamma / n $ between 0.07 and 0.04 (corresponding to $\eta$ between $4.0 \times 10^{17} \ \textrm{Pa s}$ and $6.9 \times 10^{17} \ \textrm{Pa s}$) from the black to the blue curve. For the top right panel, $\gamma / n $ is between 0.0059 and 0.0058 (corresponding to $\eta$ between $4.9 \times 10^{18} \ \textrm{Pa s}$ and $5.0 \times 10^{18} \ \textrm{Pa s}$). In the bottom left panel, $\gamma / n$ varies between 0.30 and 0.15 (corresponding to $\eta$ between $9.7 \times 10^{16} \ \textrm{Pa s}$ and $1.9 \times 10^{17} \ \textrm{Pa s}$). On the bottom right panel $\gamma / n$ is between 0.07 and 0.05 (corresponding to $\eta$ between $4.1 \times 10^{17} \ \textrm{Pa s}$ and $5.8 \times 10^{17} \ \textrm{Pa s}$).}
\label{fig4}
\end{figure}

 Fig.\,\ref{fig4} shows the rotational evolution of a non-rigid homogeneous Mercury considering two values of eccentricity and initial values of rotation rate for different values of $\gamma$. We can see that the final equilibrium value of the rotation rate depends on all those parameters. Thus, the value of $\gamma$ must be in a range such that the current rotational configuration is obtained considering the eccentricity variations of the body. There are two conditions that must be taken into account. Firstly, the relaxation factor must be such that the body does not escape the 3/2 spin-orbit resonance when its eccentricity reaches the minimum value. Also, the body may have passed the 2/1 spin-orbit resonance without remaining captured into it even when the minimum eccentricity was reached. In the case of Mercury, studies on its eccentricity history for the last 200 Myr have shown that the eccentricity oscillated between 0.1 and 0.3 (see Laskar 1996). For $e=0.1$ and by following the two criteria established just above, we obtain $5.8 \times 10^{-3} \leq \gamma / n \leq 5.8 \times 10^{-2} $, which corresponds to $ 4.8 \times 10^{-9} \ s^{-1} \leq \gamma \leq 4.8 \times 10^{-8} \ s^{-1} $ (using $n=8.2677 \times 10^{-7} \ \textrm{s}^{-1}$ as it is the case of Mercury's orbit).

\begin{figure}[!h]
\centering
\includegraphics[height=250pt,width=360pt]{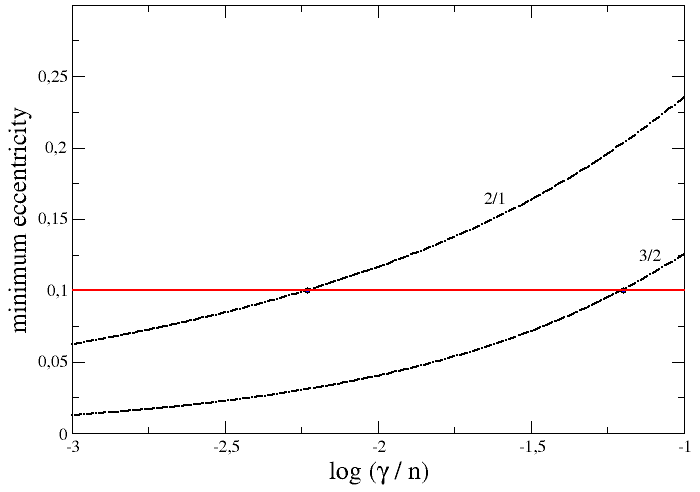}
\caption{Minimum possible eccentricity for two spin-orbit resonances as a function of $\log (\gamma / n)$. The red line shows the minimum eccentricity for Mercury and, thus, establishes the upper and lower boundary values for its $\gamma / n$ (see also Fig. 5 of Correia et al. 2014).}
\label{fig6}
\end{figure}

\par Fig.\,\ref{fig6} summarizes our results for the relaxation factor of the non-rigid body whose final rotational configuration is the 3/2 resonance. We performed numerical integrations of Eqs.\,(\ref{deltadot}) - (\ref{omegadot2}) for the body assuming, initially, that $\Omega / n = 2.2$ for a range of $\gamma / n$ between $10^{-1}$ and $10^{-3}$. We see that the only necessary information to determinate the boundary values for the relaxation factor is the minimum eccentricity of the body. The red line in the figure represents the minimum eccentricity for Mercury in the past 200 Myr. The intersection of the two black curves and the red line determinates the boundary values for Mercury's $\gamma / n$.  It is important to emphasize that, for Mercury, the rotation could have been captured in other spin-orbit resonances before the 3/2 resonant configuration was reached. If we assume, for instance, a previous fast-rotating Mercury with $\Omega / n > 2$ and $e > 0.1$, the rotation could have been temporarily trapped in the 2/1 spin-orbit resonance, provided $\gamma$ was closer to the inferior boundary value (see red line in Fig.\,\ref{fig6}). However, when Mercury reached the minimum eccentricity, the rotation would have escaped the 2/1 resonance and evolved to the 3/2 resonance.

\section{Time variation of the shape and orientation of the figure}
\label{sec:mercury}

\par
\begin{figure}
\centering
\includegraphics[height=300pt,width=400pt]{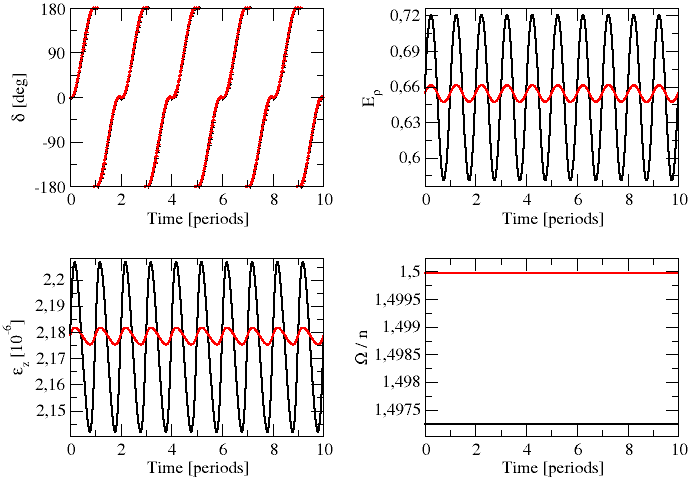}
\caption{Deviation angle ($\delta$), normalized equatorial prolateness ($E_{\rho}$), polar oblateness ($\mathcal{E} _z$),  and mean value of the rotation rate as a function of the time. The black curve corresponds to $\gamma = 4.8 \times 10^{-8}$ s$^{-1}$ and the red curve to $\gamma = 4.8 \times 10^{-9}$ s$^{-1}$. In both cases we adopted $e=0.2056$.}
\label{fig7}
\end{figure}

 Fig.\,\ref{fig7} shows the time variation of the shape and orientation of a non-rigid homogeneous Mercury's figure and the mean value of the rotation rate for the boundary values of $\gamma$ established in the previous section, based on its current rotational configuration. The mean value of the rotation is larger for the smallest value of $\gamma$. Also, the amplitudes of the oscillations of $E_{\rho}$ and $\mathcal{E} _z$ are larger for the highest value of $\gamma$.

	\begin{table}[h!]
  \begin{center}
    \caption{Values of the equatorial prolateness and polar oblateness of a homogeneous non-rigid Mercury in the 3/2 spin-orbit resonance, for $e=0.2056$.}
    \begin{tabular}{ccc} % <-- Alignments: 1st column left, 2nd middle and 3rd right, with vertical lines in between
      \hline
      \textbf{Reference} & \textbf{Equatorial prolateness} & \textbf{Polar oblateness}  \\ 
      \hline
      
      Static case (Eqs.\,\ref{nova1} and \ref{nova2}). Mean values & $1.697 \times 10^{-6}$ & $2.966 \times 10^{-6}$ \\
      Creep tide (cf. Fig.\,\ref{fig7}). Mean values & $1.11 \times 10^{-6}$ & $2.18 \times 10^{-6}$ \\
      Darwin-Kaula model (Matsuyama and Nimmo 2009) & $1.1 \times 10^{-6}$ & $2.2 \times 10^{-6}$ \\
      MESSENGER data (Perry et al. 2015) & $5.45 \times 10^{-4}$ & $9.61 \times 10^{-4}$ \\ \hline
      
    \end{tabular} 
    \label{table1}
  \end{center}
\end{table}

\par Table\,\ref{table1} shows some results of Mercury's polar oblateness and equatorial prolateness, namely: the flattenings mean values of the hydrostatic case (Eqs.\,\ref{nova1} and \ref{nova2}, corresponding to $\gamma \rightarrow \infty$), the flattenings mean values of the hydrodynamic equilibrium case of the creep tide theory (cf. Fig.\,\ref{fig7}), the results obtained by converting the values of $J_2$ and $C_{22}$ obtained by Matsuyama and Nimmo (2009; Eq. (43)) using the Darwin-Kaula model in the case of a homogeneous fluid Mercury ($k_2^{T*} = 3/2$), and the values obtained by converting the values of $J_2$ and $C_{22}$ obtained from MESSENGER's measurements (Perry et al. 2015).

\par Because of the rotation of $\delta$ and its periodicity, it is useful to characterize the figure parameters by their values at a well defined moment, namely the time of perihelion passage.

\begin{figure}
\centering
\includegraphics[height=350pt,width=470pt]{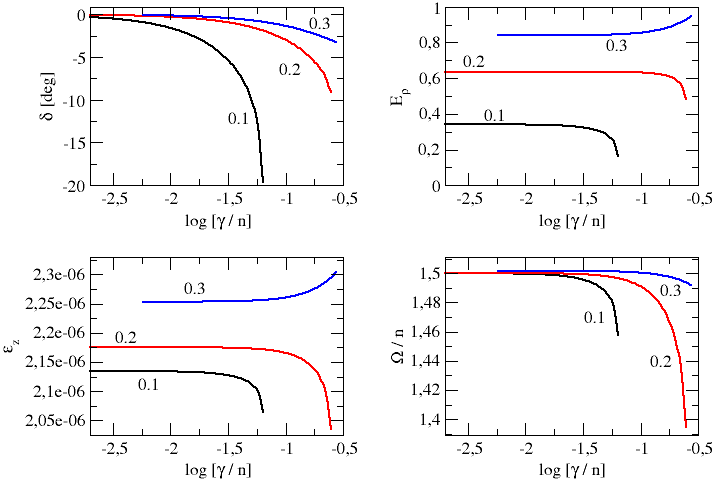}
\caption{Values of the deviation angle ($\delta$), normalized equatorial prolateness ($E_{\rho}$), polar oblateness ($\mathcal{E} _z$),  and mean value of the rotation rate in the periapsis of Mercury's orbit for three values of eccentricities as a function of $\log (\gamma / n)$. We have $e=0.1$, $e=0.2$ and $e=0.3$ on the black, red and blue curves, respectively.}
\label{fig9}
\end{figure} 
\par In Fig.\,\ref{fig9} we can see the values of the equatorial prolateness (in units of $\bar{\epsilon} _{\rho}$), polar oblateness, rotation rate and deviation angle at the perihelion of the orbit for a range of values of $\gamma$. For smaller $\gamma$, the rotation rate is closer to the exact 3/2 spin-orbit resonance, and $\delta$ is close to zero. As $\gamma$ increases, $\delta$ increases significantly until approximately 20$^\circ$ in the case of $e=0.1$ (see black curve on the top left panel of Fig.\,\ref{fig9}). Also, the value of $\Omega / n$ becomes significantly lower than 1.5 for higher values of $\gamma$.
\par Since we have shown a method to determinate the boundary values for $\gamma$ based on the final rotational configuration of the body, and applied such method to Mercury, we can now compare our results to the observed data for Mercury. Perry et al. (2015) give the values of Mercury's principal axes and deviation angle, based on MESSENGER's data. Our range of values for $\delta$ at the perihelion of Mercury's orbit encompasses the values given by Perry et al. However, for $E_{\rho}$ and $\mathcal{E} _z$, the creep tide theory predicts a value two orders of magnitude smaller than the value reported by Perry et al. There are no values of $\gamma$ for which Perry's values for $E_{\rho}$ and $\mathcal{E} _z$ could be obtained by considering tidal interactions alone. In addition, this discrepancy between the results is independent of the tidal model used. The observed values for Mercury's $E_{\rho}$ and $\mathcal{E} _z$ cannot be obtained by any tidal theory for Mercury's figure evolution, assuming hydrodynamic equilibrium.
\par Currently, it is well-known that Mercury's internal structure is not homogeneous. One attempt to solve the problem concerning the differences in $E_{\rho}$ and $\mathcal{E} _z$ is the extension of the creep tide theory for the case of a differentiated Mercury. We performed numerical integrations for a differentiated Mercury with three layers considering an approximation of the model for Mercury's interior structure proposed in Steinbr\"ugge et al. (2018). The value for the crust's relaxation factor that we have from such interior structure model is four orders of magnitude smaller than the lower boundary value that we determined for a homogeneous Mercury. However, the results taking into account Mercury's differentiated structure yields approximately the same order of magnitude for $E_{\rho}$ and $\mathcal{E} _z$ when compared to the homogeneous model (the values for the differentiated case are, in fact, smaller than the homogeneous ones). Moreover, a model taking into account the differentiation of Mercury and the lower viscosity of the fluid core of the planet should be responsible for a tidal deformation of the order of the static tide of the homogeneous model (Steinbrugge et al. 2018), thus creating a tidal torque efficient to drive the system to the capture into the 3/2 resonance. To study Mercury's rotational evolution with the current observed values for $E_{\rho}$ and $\mathcal{E} _z$, we would need to consider that there are permanent components for both the equatorial prolateness and polar oblateness coefficients (see e.g Bartuccelli et al. 2017). 

\section{Conclusion}
\label{sec:conclusion}

\par We have used the new version of the creep tide theory (Folonier et al. 2018) to study the rotational evolution of a homogeneous non-rigid body for various rotational configurations. We considered a homogeneous viscous body which is under the action of the centrifugal potential from its own rotation and the gravitational attraction of a companion. No permanent equatorial asymmetry is assumed.
\par The creep tide theory allows us to obtain the time evolution of the body's figure as well as its rotation. The differential equations to be solved in this new version of the creep tide theory are significantly simpler than the equations of the previous formulation of the theory (Ferraz-Mello 2012, 2013, 2015a). Moreover, in the new formulation of the creep tide theory, we do not assume a constant rotation rate in the early steps of the development of the theory as a working hypothesis, as it was previously done. It is worth noting that the equations of the new formulation of the creep tide theory are virtually equivalent to the equations of the Maxwell viscoelastic model studied by Correia et al. (2014). In fact, the two models become equivalent when the elastic terms are neglected (see Ferraz-Mello 2015b).
\par The dynamics of capture of a non-rigid Mercury in the 3/2 resonance was revisited. Studies of the dynamics of capture in spin-orbit resonances have been thoroughly discussed before by Makarov and Efroimsky (2013), Correia et al. (2014) and Ferraz-Mello (2014, 2015a) and much of their findings are repeated in this paper for the sake of completeness. It was shown that spin-orbit resonances ensue for bodies with a low value of the relaxation factor, and the number of spin-orbit resonances in which an initially fast rotating body may be captured increases with the eccentricity. Analytical expressions for the minimum eccentricity for which the possibility of capture in the 3/2 and 2/1 spin-orbit resonances exists were presented, their results being in excellent agreement with numerical experiments. For the comparison with the numerical experiments, we considered several numerical integrations of the equations for the time evolution of the ellipsoid's shape and the rotational evolution of the body, where the initial rotation rate was set very close to the exact values of the spin-orbit resonances (e.g $\Omega = 3n/2$ for the 3/2 spin-orbit resonance). In each numerical integration, we considered a value for the relaxation factor sufficiently close to the limit value for such resonance to exist.
\par Additionally to the study of the capture in spin-orbit resonances, we also studied the figure evolution of the body when it is trapped in the 3/2 and 2/1 spin-orbit resonances. It was shown that the tidal lag $(\delta)$ can either circulate or oscillate around 0. For instance, in the case of the 3/2 spin-orbit resonance, the angle $\varphi _B = \delta + \varphi$ circulates with the same period of the rotation, for low values of $\gamma$. For big values of $\gamma$, $\delta$ oscillates around 0, thus the bulge oscillates around the line oriented towards the companion. For the 2/1 spin-orbit resonance, it was shown that, for sufficiently low values of $\gamma$, the period of circulation of $\varphi _B$ can be either half or 2/3 of the orbital period, the former corresponding to the behavior of a rigid body. When $\gamma$ is increased, $\delta$ oscillates around 0, and $\varphi _B$ circulates with the period of the orbital motion.
\par In what concerns the application of the creep tide theory to the determination of the relaxation factor of a non-rigid Mercury based on its current spin-orbit resonant state, it was shown that the capture in the 3/2 resonance constrains the relaxation factor to be in the interval given by $ 4.8 \times 10^{-9} \ s^{-1} \leq \gamma \leq 4.8 \times 10^{-8} \ s^{-1} $ (corresponding to $5.1 \times 10^{17} \ \textrm{Pa s} \leq \eta \leq 5.0 \times 10^{18} \ \textrm{Pa s}$) when we assume, as working hypotheses, that (i) Mercury's eccentricity has never been below $e=0.1$, (ii) its rotation rate was initially faster than the current value, and (iii) no permanent components of the flattenings existed at the time of capture in the 3/2 spin-orbit resonance. These estimated values for $\eta$ are much smaller than the values used by Steinbrugge et al. (2018), who assumed that $\eta = 10^{23} \ \textrm{Pa s}$ (which corresponds to $\gamma$ of the order $10^{-13} \ \textrm{s}^{-1}$) for Mercury's crust. Some numerical experiments were performed, supposing a differentiated structure for Mercury (based on the model of Folonier and Ferraz-Mello 2019), using the internal structure model proposed by Steinbrugge et al. (2018). We verified that Mercury's 3/2 spin-orbit resonant state can be maintained for the value of Mercury's crust viscosity of Steinbrugge et al. (2018), if the viscosity of the liquid core is small. These preliminary experiments have shown that the rotation of the core may be rapidly driven to the 3/2 spin-orbit resonance when we consider a linear friction acting between the crust and the core.
\par Finally, the results for the values of the equatorial prolateness and polar oblateness of a non-rigid homogeneous Mercury were compared to the data obtained from the MESSENGER mission. The predicted flattenings, which are at most $\mathcal{E}_{\rho} = 1.1 \times 10^{-6}$ and $\mathcal{E}_{z} = 2.2 \times 10^{-6}$ according to the tidal theories, were compared to the values obtained from MESSENGER observations, which give $\mathcal{E}_{\rho} = 5.5 \times 10^{-4}$ and $\mathcal{E}_{z} = 9.6 \times 10^{-4}$ (see Perry et al. 2015). The observed equatorial prolateness and polar oblateness coefficients are two orders of magnitude bigger than the values predicted by the tidal theories. This discrepancy between the results for the flattenings holds for any tidal theory based on hydrodynamic equilibrium. We may add that, in the case of a model considering a differentiated structure for Mercury, the values of the crustal flattenings are even smaller than the values of the flattenings of the homogeneous model. These facts suggest the existence of a fossil component, responsible for Mercury's current shape.

\section*{Acknowledgements}

We thank the two referees for the fruitful discussions about tides and equilibrium figures, which led to an improvement of this paper. This investigation is funded by the National Research Council, CNPq, grant 302742/2015-8 and by FAPESP, grants 2016/20189-9 and 2017/25224-0. This investigation is part of the thematic project FAPESP 2016/13750-6.

\section*{Additional Information}

\par Competing Interests: The authors declare no competing interests.

\end{document}